\documentclass[12pt]{article}
\usepackage{amssymb}
\usepackage{amsmath}
\usepackage{amsthm}
\usepackage{cancel}
\usepackage{slashed}
\usepackage{fullpage}
\usepackage{color}
\usepackage{braket}
\usepackage{graphicx}
\usepackage[small]{subfigure}
\usepackage{multirow}
\usepackage{verbatim}
\usepackage{cite}
\usepackage{bigints}
\usepackage{simplewick}
\usepackage{authblk}
\usepackage{hyperref}
\usepackage{empheq}
\usepackage{tikz-feynman}
\usepackage{tikz}
\usepackage{float}
\hypersetup{
    linktocpage,
     colorlinks,
     citecolor=darkgreen,
     linkcolor= darkgreen,
     urlcolor=darkgreen
}

\definecolor{darkred}{rgb}{0.5,0.0,0.0}
\definecolor{darkblue}{rgb}{0.0,0.0,0.9}
\definecolor{darkerblue}{rgb}{0.0,0.0,0.5}
\definecolor{darkgreen}{rgb}{0.0,0.5,0.0}
\definecolor{black}{rgb}{0.0,0.0,0.0}
\definecolor{brown}{rgb}{0.6,0.4,0.2}

\def\be{\begin{equation}}
\def\ee{\end{equation}}

\newcommand{\fd}[2]{\parbox{#1}{\includegraphics[width=#1]{figs/#2}}}

\numberwithin{equation}{section}

\newcommand{\RNum}[1]{\uppercase\expandafter{\romannumeral #1\relax}} 	
\title{Gauge invariance of the vector meson mass in the Coleman-Weinberg model}

\author{Haojie Shen$^{1}$\thanks{haojieshen@mail.ecust.edu.cn}}
\author{Yu Cheng$^{1,2}$\thanks{chengyu@mail.ecust.edu.cn}}
\author{Wei Liao$^1$\thanks{liaow@ecust.edu.cn}}

\affil{{\vskip 0.2cm
		${}^{1}$Institute of Modern Physics,  School of Sciences,\\
		East China University of Science and Technology, 130 Meilong Road, Shanghai 200237, P. R. China}}
\affil{{\vskip 0.1cm ${}^{2}$Tsung-Dao Lee Institute, and School of Physics and Astronomy, \\
Shanghai Jiao Tong University, Shanghai 200240, P. R. China}}

\begin{document} 
\maketitle

\begin{abstract}
We revisit the problem of the gauge invariance in the Coleman-Weinberg model in which a $U(1)$ gauge symmetry is
driven spontaneously broken by radiative corrections. It was noticed in previous work that masses in this model are
not gauge invariant at one-loop order. In our analysis, we use the dressed propagators of scalars which includes a resummation 
of the one-loop self-energy correction to the tree-level propagator. We calculate the one-loop self-energy correction to 
the vector meson using these dressed propagators. We find that the pole mass of the vector meson  calculated  using
 the dressed propagator is gauge invariant at the vacuum determined using the effective potential calculated with 
a resummation of daisy diagrams. 
\end{abstract}

\section{Introduction}
\qquad 
One of the subtle problems in quantum field theory(QFT) is the gauge invariance of physical quantities.
It is generally believed that physical quantities such as the S-matrix elements, the physical masses, the energy density of a physical state, 
the decay rate of a false vacuum, etc.,
are gauge invariant although the quantum field theory can be quantized in an arbitrary gauge. One quantity often encountered in QFT is the effective potential which is usually interpreted as the energy density
just as the potential is usually interpreted as the energy density.

However, it was found that the effective potential in the Coleman-Weinberg(CW) model~\cite{Coleman:1973jx}, 
which is a model of massless scalar quantum electrodynamics(QED) with symmetry broken by radiative corrections,
is not gauge invariant~\cite{Jackiw:1974cv}.
This has raised concerns about how the gauge invariance is achieved in physical quantities in scalar QED and in general scalar gauge theories
and what is the meaning of the effective potential.
Many attempts have been made to clarify this problem in different models for various quantities
\cite{Nielsen:1975fs, Kang:1974yj, Nielsen:1987ht, Johnston:1986cp, Bazeia:1988, Larsen:1989xq, Quiros:1992ez, Tye:1996au, Wainwright:2011qy,Garny:2012cg,
Andreassen:2014eha, Metaxas:1995ab,Plascencia:2015pga, Espinosa:2016uaw, DiLuzio:2014bua,Andreassen:2014gha,Chigusa:2018uuj,Chigusa:2017dux,
Akil:2020prj,Metaxas:2020atd}, e.g. for the decay rate of the false vacuum~\cite{Metaxas:1995ab, Plascencia:2015pga,Chigusa:2018uuj,Chigusa:2017dux}.

One interesting observation is that although the effective potential is in general not gauge invariant, 
but the minimum value of the effective potential turns out to be gauge invariant\cite{Nielsen:1975fs}.
So the minimum value of the effective potential is indeed a physical quantity.
Another interesting observation is that 
the pole masses in the CW model are not gauge invariant at one-loop order,
but the ratio of the pole masses of  the scalar meson and the vector meson are gauge invariant~\cite{Kang:1974yj}.
These results are encouraging but are still far from a clear clarification.
In particular, the gauge dependence of the pole masses in CW model is apparently not satisfactory. It is a puzzling result. 
The gauge independence of the minimum of effective potential is also not easy to see clearly although the argument
given in \cite{Nielsen:1975fs} looks straightforward. 

It was noticed in \cite{Nielsen:1975fs,Nielsen:1987ht} that an infinite series of diagrams, the so-called daisy diagrams, can give contributions
of the same order of the gauge-dependent part in the one-loop  contribution to the effective potential and these diagrams should be taken into
account in a consistent analysis~\cite{Bazeia:1988}.  In was shown that these contributions can also be evaluated by summing the daisy diagrams 
into the dressed propagators of scalar fields~\cite{Johnston:1986cp}. 
A recent publication~\cite{Andreassen:2014eha} shows that after a careful calculation of the daisy diagrams 
the gauge invariance of the minimum of the effective potential can be achieved. 
Actually, daisy re-summation also plays important role in solving other problems, e.g. the gauge dependence problem in finite-temperature field theory \cite{Quiros:1992ez,Wainwright:2011qy,Garny:2012cg} and the infrared problem caused by the massless Goldstone boson\cite{Martin:2014bca,Elias-Miro:2014pca}.  In view of these recent advancements, it is natural to expect that this idea can help to solve other remaining problems, e.g. the gauge dependence
of the physical mass in the CW model.

%

In this article, we re-analyze the radiative correction to the pole mass of the vector meson in the CW model at one-loop order.
We show that the pole mass of the vector meson in CW model is gauge invariant at one-loop order
after including the daisy re-summation effects in the propagator and in the effective potential. 
This result is the novel contribution of the present article.
This was argued in \cite{Andreassen:2014eha} to happen, but a concrete calculation has not been done yet.
Since this problem has been a  puzzle for a long time,  it's worth making a detailed presentation although the content may look pedagogical.

The article is organized as follows. In the next section, we quickly review the CW model and the effective potential
after including the resummation effect. We obtain an expression of vacuum using this effective potential.
Then we address the resummation effects in the propagators and 
calculate the one-loop self-energy of the vector meson when including these effects.
We find that the gauge dependence in the  pole mass of the vector meson is completely cancelled at one-loop order after including 
all these resummation effects. We give some details of our calculation in the Appendix.

\section{The effective potential and the vacuum}

\qquad 
The CW model is a model with a massless scalar field coupled  with a $U(1)$ gauge field.
The Lagrangian of the CW model can be written as follows 
\begin{equation}
\mathcal{L}=-\frac{1}{4} F_{\mu \nu}^{2}+\frac{1}{2}\left(\partial_{\mu} \phi_{1}-e A_{\mu} \phi_{2}\right)^{2}+\frac{1}{2}\left(\partial_{\mu} \phi_{2}+e A_{\mu} \phi_{1}\right)^{2}-V(\phi)+\mathcal{L}_{\mathrm{GF}}
\end{equation}
where $\phi_1$ and $\phi_2$ are two components of the complex scalar field $\phi$,  $A_\mu$ is the $U(1)$ gauge field and
$F_{\mu\nu}$ the field strength, and the potential $V$ is
\begin{equation}
V(\phi)=\frac{\lambda}{24} \phi^{4}  \label{Vtree}
\end{equation}
with $\phi^{2} \equiv \phi_{1}^{2}+\phi_{2}^{2}$. 
$\mathcal{L}_{\mathrm{GF}}=-\frac{1}{2 \xi}\left(\partial_{\mu} A^{\mu}\right)^{2}$ is the gauge-fixing term.
The tree level potential of the CW model has a minimum at $\phi = 0$,  where the original symmetry is  not broken. 
However, quantum correction can give rise to an effective potential which can drive the minimum away from the origin,  as pointed out 
in \cite{Coleman:1973jx}. This leads to symmetry breaking driven by radiative corrections.

The effective potential can be obtained using the background field method~\cite{Jackiw:1974cv}.
The effective potential including the leading quantum correction is obtained as
\begin{equation}
	V_{e^4}(\hat{\phi})= \frac{\lambda}{24} \hat{\phi}^4 + \frac{\hbar e^4}{16\pi^2} \hat{\phi}^4  \left(-\frac{5}{8} 
	+ \frac{3}{2} \ln \frac{e \hat{\phi}}{\mu} \right),  \label{Ve4}
\end{equation}
where $\hat{\phi}$ is a background field taken as the value of $\phi_1$, i.e. $\phi_1=\hat{\phi}$ and $\phi_2=0$,
and $\mu$ is the renormalization scale.
The effective potential in Eq. (\ref{Ve4}) has been renormalized in the modified minimal subtraction($\overline{MS}$) scheme.
We can see  in Eq. (\ref{Ve4}) that if $\lambda \sim \frac{\hbar}{16 \pi^{2}} e^{4}$, 
the potential can have a minimum for $\phi \neq 0$ and then the symmetry is broken spontaneously driven by quantum corrections.

The effective potential in Eq.  (\ref{Ve4}) is gauge invariant. However, there are other terms in the one-loop quantum correction.
In particular, there is a term of order $\lambda e^2$ which depends on the gauge-fixing parameter $\xi$, as shown in \cite{Jackiw:1974cv}.
Since $\lambda\sim e^4$ in this model, this gauge dependent term is of higher order,  i.e. of order $e^6$.
As a consequence, the masses obtained using this effective potential are not gauge invariant.
It was shown in~\cite{Kang:1974yj} that the scalar-to-vector mass ratio is gauge independent if including
two loop corrections in the effective potential, but the masses of the scalar meson and the vector meson are still gauge dependent.

\begin{figure}[H]
	\centering
	\includegraphics[width=0.45\linewidth]{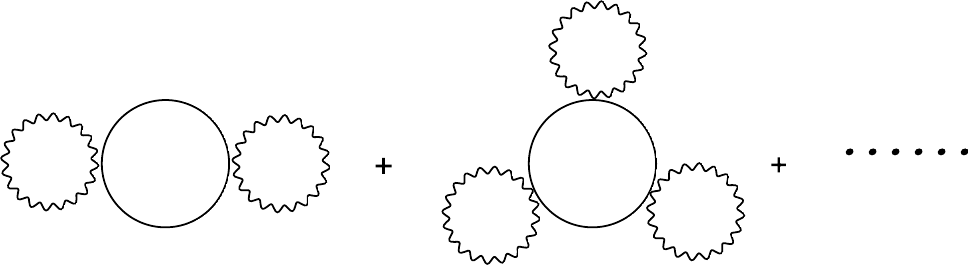}
	\caption{High-loop graphs which contribute terms proportional to $\frac{e^{10}}{\lambda}$ to the effective potential.}
	\label{fig:daisy}
\end{figure}

It was suggested in \cite{Nielsen:1975fs,Nielsen:1987ht} that all the higher loop contributions given in
Fig. \ref{fig:daisy}, the so-called daisy diagrams, should be included when calculating the effective potential.
This is because these diagrams would give contribution of order $e^{10}/\lambda$ which is of order $e^6$ for $\lambda \sim e^4$.
A re-summation of the contributions of these diagrams have been done in ~\cite{Andreassen:2014eha}
and an effective potential is obtained as follows
\begin{equation}
	V(\hat{\phi})=V_{e^4}(\hat{\phi})+V_{e^6}(\hat{\phi})  \label{Veff}
\end{equation}
where $V_{e^4}$ is given in Eq.(\ref{Ve4}) and $V_{e^6}$ in the $\overline{MS}$ scheme is \cite{Andreassen:2014eha}
\begin{small}
	\begin{equation}
		\begin{aligned}
			&V_{e^6}(\hat{\phi})= \frac{\hbar e^{2} \lambda}{16 \pi^{2}} \hat{\phi}^{4}\left(\frac{\xi}{8}-\frac{\xi}{24} \ln \frac{e^{2} \lambda \xi \hat{\phi}^{4}}{6 \mu^{4}}\right) \\
			&+\frac{\hbar^{2} e^{6}}{\left(16 \pi^{2}\right)^{2}} \hat{\phi}^{4}\left[(10-6 \xi) \ln ^{2} \frac{e \phi}{\mu}+\left(-\frac{62}{3}+4 \xi-\frac{3}{2} \xi \ln \frac{\lambda \xi}{6 e^{2}}\right) \ln \frac{e \hat{\phi}}{\mu}+\xi\left(-\frac{1}{2}+\frac{1}{4} \ln \frac{\lambda \xi}{6 e^{2}}\right)+\frac{71}{6}\right] \\
			&+ \frac{\hbar e^{2} \lambda}{16 \pi^{2}}\hat{\phi}^{4}\left(-\frac{\xi}{24}\right)\left[\frac{\widehat{\lambda}(\hat{\phi})}{\lambda}+\left(1-\frac{\widehat{\lambda}(\hat{\phi})}{\lambda}\right) \ln \left(1-\frac{\widehat{\lambda}(\hat{\phi})}{\lambda}\right)\right]
		\end{aligned}\label{V_e6}
	\end{equation}
\end{small}
where
\begin{align}
	&\widehat{\lambda}(\hat{\phi}) \equiv \frac{ \hbar e^{4}}{16 \pi^{2}}\left(6-36 \ln \frac{e \hat{\phi}}{\mu}\right) . \label{hatlambda}
\end{align}
The second part in Eq.(\ref{V_e6})  comes from two loop contributions and 
the third part in Eq.(\ref{V_e6}) is the contribution of the resummation of daisy diagrams in Fig. \ref{fig:daisy}.
Using this effective potential, it was shown explictly that the effective potential at its minimum is gauge invariant to order $e^6$~\cite{Andreassen:2014eha}.

The vacuum expectation value, $\hat{\phi}=\upsilon$, can be obtained from extremum condition
\begin{equation}
	\left.\frac{\partial V(\hat{\phi})}{\partial\hat{\phi}}\right|_{\hat{\phi}=\upsilon}=0.\label{define_vacuum1}
\end{equation}
Since $V(\hat{\phi})$ is not gauge invariant, $\upsilon$ is not gauge invariant either.
However, it can be found that $\upsilon$ is gauge invariant at the leading order and is gauge dependent at order $e^2$.
So we can express $\upsilon$ as follows
\begin{equation}
	\upsilon=\upsilon_0+\upsilon_{1}+\cdots, \label{vacuum}
\end{equation}
where $v_0$ is the value of $\upsilon$ at leading order, $\upsilon_{1}$ the correction at order $e^2$.
 Other higher order corrections have been neglected.
$\upsilon_{0}$ can be obtained by considering the extremum condition of $V_{e^4}$, i.e.
\begin{equation}
	\left.\frac{\partial V_{e^4}(\hat{\phi})}{\partial\hat{\phi}}\right|_{\hat{\phi}=\upsilon_0}=0.\label{define_vacuum2}
\end{equation}
It gives
\begin{equation}
	\lambda-\widehat{\lambda}(\upsilon_0)=0,\label{define_vacuum2a} 
\end{equation}
where $\widehat{\lambda}$ is given in Eq. (\ref{hatlambda}). $\upsilon_0$ determined using Eq. (\ref{define_vacuum2a}) is gauge invariant 
because the expression of $\widehat{\lambda}$ in Eq. (\ref{hatlambda}) does not depend on the gauge-fixing parameter $\xi$.

The coupling constants $\lambda$ and $e$ in Eq. (\ref{define_vacuum2a} ) should be understood as
renormalized at the scale $\mu$. Using the $\beta$ functions of $\lambda$ and $e$ at the leading order~\cite{Andreassen:2014eha}
\begin{equation}
\mu \frac{d}{d \mu} \lambda=\frac{9\hbar e^4}{4\pi^2},  ~~\mu \frac{d}{d \mu} e=\frac{\hbar e^3}{48\pi^2},
\end{equation}
it is straightforward to show that $v_0$ determined in Eq.  (\ref{define_vacuum2a} )  does not depend on 
the renormalized scale $\mu$ at the leading order, i.e. at order $e^4$. So it is convenient to choose a fixed energy scale $\mu_X$ 
for which the running coupling constants satisfy the following relation~\cite{Andreassen:2014eha} 
\begin{equation}
\lambda(\mu_X)=\frac{ \hbar }{16 \pi^{2}} e^4(\mu_X) \{6-36 \ln[ e(\mu_X)] \}. \label{muX}
\end{equation} 
We can choose $\mu =\mu_X$ in Eq. (\ref{define_vacuum2a} ). Then it is easy to see that 
\begin{equation}
v_0=\mu_X.  \label{v0}
\end{equation}
Since $v_0$ does not depend on $\mu$, $v_0$ found in this way is the same for arbitrary value of  $\mu$.

Now we can obtain an expression of $\upsilon_{1}$.  Notice  that  Eq. (\ref{define_vacuum2}) means
 \begin{equation}
	\left.\frac{\partial V_{e^4}(\hat{\phi})}{\partial\hat{\phi}}\right|_{\hat{\phi}=\upsilon_0+\upsilon_1}=
	\left.\frac{\partial^2 V_{e^4}(\hat{\phi})}{\partial\hat{\phi}^2}\right|_{\hat{\phi}=\upsilon_0} \upsilon_1
	+O(\upsilon^2_1)+\cdots .\label{define_vacuum3}
\end{equation}
Keeping term linear in $\upsilon_1$ in Eq. (\ref{define_vacuum3}),  we can write Eq. (\ref{define_vacuum1}) as
\begin{equation}
\left.\frac{\partial^2 V_{e^4}(\hat{\phi})}{\partial\hat{\phi}^2}\right|_{\hat{\phi}=\upsilon_0} \upsilon_1
+\left.\frac{\partial V_{e^6}(\hat{\phi})}{\partial\hat{\phi}}\right|_{\hat{\phi}=\upsilon_0+\upsilon_1}=0. \label{define_vacuum}
\end{equation}
Plugging Eq. (\ref{V_e6}) into Eq. (\ref{define_vacuum})  and keeping terms of the leading order, we obtain
\begin{eqnarray}
\upsilon_1=&&\frac{\hbar e^{2}}{16 \pi^{2}} \frac{\mu_{X}}{2}\bigg[ \big( -\xi-\frac{80}{9}\big)-\frac{40}{3} \ln^{2} e
+\frac{94}{9} \ln e^{2} \nonumber \\
	  &&+\frac{1}{2} \xi\bigg(\ln \frac{\lambda\xi e^2}{6}+\ln \big(1-\frac{\widehat{\lambda}(\upsilon)}{\lambda}\big)\bigg)  \bigg].
	\label{v1_vacuum_final}
\end{eqnarray}
In Eq. (\ref{v1_vacuum_final}), the $\ln \big(1-\frac{\widehat{\lambda}(v)}{\lambda}\big)$ have been kept as a function of 
$\upsilon=\mu_X +\upsilon_1$ rather than of $\mu_X$, 
because this logarithm diverges at $\hat{\phi}=\mu_X$.
So Eq. (\ref{v1_vacuum_final}) can be understood as an iterative expression of $\upsilon_1$.
We can see in Eq. (\ref{v1_vacuum_final}) that $\upsilon_1$ is indeed of order $e^2$.
This expression of $\upsilon_1$ will be used later when studying the gauge invariance of the vector meson mass.

\section{The dressed propagator}

Using the background field $\hat{\phi}$, the effective propagators of scalars at tree-level are~\cite{Kang:1974yj}
\begin{equation}
	\begin{aligned}
		1~\fd{2cm}{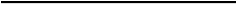}~1 ~~=~~ D^0_{11}(k) &= \frac{i}{k^2-\frac{\lambda}{2}\hat{ \phi}^2}, \\
		2~\fd{2cm}{scalarline}~2 ~~=~~ D^0_{22}(k) &=  \frac{i(k^2 - e^2 \xi \hat{\phi}^2)}{k^{4}-\frac{\lambda}{6}\hat{ \phi}^{2}\left( k^2-\xi e^2\hat{ \phi}^{2}\right) }. \\
	\end{aligned}  \label{D-0}
\end{equation}

\begin{figure}[H]
	\centering
	\includegraphics[width=0.25\linewidth]{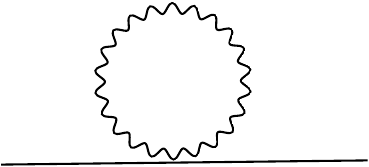}
	\caption{Transverse vector tadpole graph $A_{\mathrm{loop}}$}
	\label{fig:insert}
\end{figure}

The key point of daisy summation is that the self-energy diagram shown in Fig. \ref{fig:insert} is of the same order of tree-level contribution.
In fact, a straightforward calculation of this diagram using the dimensional regulation gives
\begin{equation}
	\begin{aligned}
			A_{\mathrm{loop}}=\frac{\hbar}{2}\int\frac{d^Dk}{(2 \pi)^{D}} 2ie^{2}g_{\mu \nu}\frac{-i}{k^{2}-e^{2} \hat{\phi}^{2}}\left(g^{\mu \nu}-\frac{k^{\mu} k^{\nu}}{k^{2}}\right)=\frac{i\hbar}{16\pi^2}e^4\hat{ \phi}^2(1-6\ln\frac{e\hat{ \phi}}{\mu}),
	\end{aligned}  \label{Aloop}
\end{equation}
where we have performed subtraction in the $\overline{MS}$ scheme and 
have eliminated the term proportional to  $\Delta_{\varepsilon}=\frac{1}{\varepsilon}-\gamma_{E}+\ln 4\pi$.
One can see clearly in Eqs. (\ref{D-0}) and (\ref{Aloop}) that the self-energy diagram in Fig. \ref{fig:insert} is of the same order of the 
tree-level mass which is $\sim \lambda \hat{\phi}^2\sim e^4 \hat{\phi}^2$.

\begin{figure}[H]
	\centering
	\includegraphics[width=0.6\linewidth]{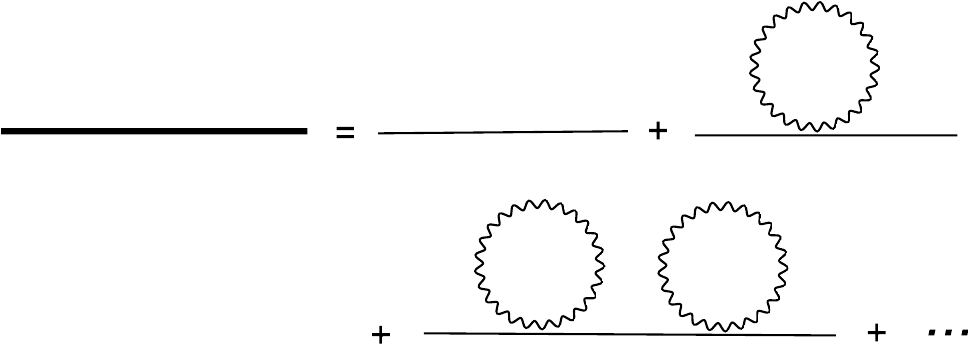}
	\caption{The dressed propagator of $\phi_2$ as a resummation of a series of vector tadpole diagrams inserted into tree-level propagators.}
	\label{fig:resummation}
\end{figure}

This suggests that we should include a geometric summation shown in Fig. \ref{fig:resummation} in calculation of 
quantum corrections~\cite{Johnston:1986cp,Larsen:1989xq} .
This summation gives  modified propagators, called dressed propagators. For example, for $\phi_2$ the summation gives 
a dressed propagator as follows
\begin{eqnarray}
D_{22} &&=D^0_{22}+D^0_{22} A_{\mathrm{loop}} D^0_{22} 
                     +D^0_{22} A_{\mathrm{loop}} D^0_{22}A_{\mathrm{loop}} D^0_{22}+\cdots \nonumber \\
   &&=\frac{i(k^2 - e^2 \xi \hat{\phi}^2)}{k^{4}- m^{2}_{2} \left( k^2-\xi e^2\hat{ \phi}^{2}\right)},
\end{eqnarray}
with
\begin{equation}
m^{2}_{2}(\hat{\phi})= \frac{\lambda}{6}\hat{ \phi}^{2}+iA_{\mathrm{loop}}=\frac{\lambda-\widehat{\lambda}}{6}\hat{\phi}^{2}.
\label{m2}
\end{equation}
where $\widehat{\lambda}$ is given in Eq. (\ref{hatlambda}).

For $\phi_1$, this geometric summation using vector tadpole would give rise to a mass shifting to $\frac{3\lambda-\widehat{\lambda}}{6} \hat{\phi}^2$.
However, this is not the correct result. There are other self-energy diagrams which can contribute at order $e^4$. 
These self-energy diagrams have been computed  in ~\cite{Kang:1974yj} as a momentum expansion around $p^2=0$.
Restoring the coupling constant in logarithm, which is omitted in \cite{Kang:1974yj}, 
the result which is of the same order of $ i (D^0_{11})^{-1}$ can be extracted out as
\begin{equation}
\Sigma_{11}=-i \frac{\hbar e^4}{16\pi^2} \hat{\phi}^2 ( 3+18  \ln \frac{e \hat{\phi}}{\mu} ~).  \label{Sigma11}
\end{equation}
Performing a summation using Eq. (\ref{Sigma11}) we can get a dressed propagator for $\phi_1$ as
\begin{eqnarray}
D_{11} &&=D^0_{11}+D^0_{11} \Sigma_{11} D^0_{11} 
                     +D^0_{11} \Sigma_{11} D^0_{11} \Sigma_{11} D^0_{11}+\cdots \nonumber \\
   &&=\frac{i}{k^{2}- m^{2}_{1} },
\end{eqnarray}
with
\begin{equation}
m^{2}_{1}(\hat{\phi})= \frac{\lambda}{2}\hat{ \phi}^{2}+i\Sigma_{11}
=\frac{\lambda}{2}\hat{\phi}^{2}+\frac{\hbar e^4}{16\pi^2} \hat{\phi}^2 ( 3+18  \ln \frac{e \hat{\phi}}{\mu} ~).
\label{m1}
\end{equation}
One can also check that other self-energy diagrams for $\phi_2$ give zero or are of higher order than the vector tadpole diagram in Fig. \ref{fig:insert}.
So the correction given in Eq. (\ref{m2}) is the total leading order one-loop correction to the dressed mass of $\phi_2$.

Because $ (D^0_{22})^{-1}$, part of the whole matrix of the inverse of the propagators for $(\phi_1,\phi_2, A_\mu)$, 
has been modified to $(D_{22})^{-1}$,  other parts of  propagators related to $D^0_{22}$ should also be modified.
This can be seen clearly when solving the propagator matrix of  $(\phi_1,\phi_2, A_\mu)$ from the matrix of 
the inverse propagators, as was done in \cite{Kang:1974yj}.
The result is a shift of the $\phi_2$ mass to $m_2^2(\hat{\phi})$ in all these related propagators.
These dressed propagators have been summarized in Appendix \ref{AppendixA}.

We note that the dressed masses obtained in Eqs. (\ref{m2}) and (\ref{m1}) are consistent with the results one can obtain using 
the effective potential~\cite{Metaxas:1995ab}. In fact, using Eqs. (\ref{Ve4}) one can easily verify
\begin{equation}
	m^{2}_{1}(\hat{\phi})=\frac{\partial^2V_{e^{4}}}{\partial\hat{\phi}^2},~~~~     
	m^{2}_{2}(\hat{\phi})=\frac{1}{\hat{\phi}}\frac{\partial V_{e^{4}}}{\partial \hat{\phi}}.
\end{equation}
So $\phi_2$ becomes massless at the vacuum at the leading order determined by $V_{e^4}$, as it should happen as a Goldstone boson.

\section{The vector meson mass}
\qquad 
Using the dressed propagators and the vertices shown in Appendix \ref{AppendixA}, we can calculate the one-loop self-energy diagrams 
of the vector meson and the radiative corrections to the vector meson mass. 
The relevant one-loop self-energy diagrams are shown in Fig. \ref{fig:vector} in  Appendix \ref{AppendixB}
and the corresponding Feynman integrals are given in Eqs. (\ref{B1-a}), (\ref{B1-b}),  (\ref{B1-c}),  (\ref{B1-d})
and (\ref{B1-e}).  The total self-energy $-i\Sigma_{\mu \nu}$ is the sum of the contributions of all these diagrams. 

Using $\Sigma_{\mu \nu}$, we can write down the inverse of propagator of the vector meson at one-loop order
\begin{equation}
	G^{-1}_{\mu \nu}\left(p^{2}\right)= i g_{\mu \nu}\left( p^{2}-e^{2} \hat{\phi}^{2}\right)+i \Sigma_{\mu \nu}\left(p^{2}\right)+p_{\mu}p_{\nu}\,\mathrm{term}, \label{green_function_0}
\end{equation}
where the first term is the tree-level contribution. 
$\Sigma_{\mu \nu}$ can also be written as 
\begin{equation}
\Sigma_{\mu \nu}(p^2)= g_{\mu \nu}\Sigma(p^2) + p_{\mu}p_{\nu} ~\mathrm {term}.  \label{self-energy}
\end{equation}
So we can rewrite Eq. (\ref{green_function_0}) as
\begin{equation}
	G^{-1}_{\mu \nu}\left(p^{2}\right)= i g_{\mu \nu}\left( p^{2}-e^{2} \hat{\phi}^{2}+\Sigma(p^2) \right)+p_{\mu}p_{\nu}\,\mathrm{term}, \label{green_function_1}
\end{equation}
The pole mass of the vector meson, $m^2_v$, is determined by the condition that the first term in
Eq. (\ref{green_function_1}) vanishes at the point $p^2=m^2_v$.

$\Sigma_{\mu \nu}$ can be evaluated using a momentum expansion around the point $p^2=0$ as in ~\cite{Kang:1974yj}.
Introducing
\begin{equation}
\delta m^{2}=- \Sigma(p^2=0),~~
z_{2}=-\left.\frac{\partial(\Sigma(p^2))}{\partial p^{2}}\right|_{p^2=0}, \label{expansion1}
\end{equation}
and neglecting terms of higher orders in $p^2$, we can write Eq. (\ref{green_function_1}) as
\begin{equation}
	\begin{aligned}
	G^{-1}_{\mu \nu}\left(p^{2}\right)=ig_{\mu \nu}\left[p^2-e^{2} \hat{\phi}^{2}-\delta m^{2}-z_{2}p^2\right]+p_{\mu}p_{\nu}\,\mathrm{term}.
	\end{aligned}\label{green_function_2}
\end{equation}
Detailed results to order $e^4$ have been given in Appendix \ref{AppendixC}.  Using results in Appendix \ref{AppendixC}, we can find
\begin{equation}
	\begin{aligned}
		\delta m^{2}=\frac{\hbar }{16 \pi^{2}}e^4 \phi^{2}\left[ \xi-\frac{5}{2} -\frac{\xi}{2} 
		\ln \left(\frac{e^{2} \hat{\phi}^{2} \xi m_{2}^{2}}{\mu^{4}}\right) +3\ln\frac{e^2\hat{ \phi}^2}{\mu^{2}} \right],
	\end{aligned} \label{deltam2}
\end{equation}
and 
\begin{equation}
	\begin{aligned}
		z_{2}=\frac{\hbar e^2}{16 \pi^{2}}\left( -\frac{41}{18}+\frac{1}{3}\ln\frac{e^2\hat{ \phi}^2}{\mu^{2}}  \right) 
	\end{aligned}, \label{zeta2}
\end{equation}
where terms proportional to $\Delta_{\varepsilon}$ have been subtracted in renormalization. 
In Eqs. (\ref{deltam2}) and (\ref{zeta2}), we have kept the coupling constants and the gauge-fixing parameter $\xi$ in the
logarithms. If restoring the tree-level mass for $\phi_1$ and $\phi_2$ and 
neglecting the coupling constants and the gauge-fixing parameter $\xi$ in the logarithms, we agree with the results in ~\cite{Kang:1974yj}.

Using Eqs. (\ref{green_function_2}),  (\ref{deltam2}) and (\ref{zeta2}), we can find the pole mass to $e^4$ as follows 
\begin{eqnarray}
m_{v}^{2} &&=e^2 \hat{\phi}^{2}+\delta m^{2}+z_{2}e^2\hat{\phi}^{2} \nonumber \\
&&=e^2\hat{\phi}^2+\frac{\hbar}{16\pi^2}e^4\hat{\phi}^2\left[ \xi-\frac{43}{9}
-\frac{\xi}{2} \ln \left(\frac{e^{2} \hat{\phi}^{2} \xi m_{2}^{2}}{\mu^{4}}\right)+\frac{10}{3}\ln \frac{e^{2} \hat{\phi}^{2}}{\mu^{2}}\right] .
\label{polemass1}
\end{eqnarray}

Now we set the scale $\mu=\mu_{X}$ and consider the case at vacuum $\hat{\phi}=v$.
Using Eq. (\ref{vacuum}) and keeping terms to order $e^4$, we find
$m_{v}^{2}(v)$  up to order $e^4$
\begin{equation}
	\begin{aligned}
		&m_{v}^{2}(\upsilon)=e^2\mu_{X}^{2}+2e^2\mu_{X}\upsilon_1\\
		&+\frac{\hbar}{16\pi^2}e^4\mu_{X}^{2}\left[ \xi-\frac{43}{9}+\frac{10}{3}\ln e^2
		-\frac{\xi}{2}\left( \ln(\frac{e^2\lambda\xi}{6})+\ln\left(1-\frac{\widehat{\lambda}(\upsilon)}{\lambda}\right)\right) \right]
	\end{aligned}\label{m_0}
\end{equation}

Inserting Eq.(\ref{v1_vacuum_final}) into Eq.(\ref{m_0}),  it is easy to see that the $\xi$ dependence is cancelled completely.
We find the pole mass of the vector meson as
\begin{equation}
	\begin{aligned}
			m_{v}^{2}
			&=e^2\mu_{X}^{2}+\frac{\hbar}{16\pi^2}e^4\mu_{X}^{2}\left( -\frac{123}{9}+\frac{124}{9}\ln e^2 -\frac{40}{3} \ln^{2} e\right)
	\end{aligned}\label{polemass1-a}
\end{equation}
It is manifestly a gauge invariant result.

So far we have used a momentum expansion around $p^2=0$ in the  evaluation of $\Sigma_{\mu \nu}(p^2)$ or $\Sigma(p^2)$.
However, there is a problem in this expansion. If expanding $\Sigma(p^2)$  around $p^2=0$, such as 
\begin{equation}
\Sigma(p^2)=\Sigma(0)-z_2 ~p^2 + z_4 ~p^4+\cdots, \label{expansion2}
\end{equation}
we hope this expansion is convergent, i.e. $|z_4 ~p^4| \ll |z_2 ~p^2|$ for $p^2\approx e^2 \hat{\phi}^2$.
However, the natural dimensional parameter appearing in the denominator is $e^2 \hat{\phi}^2$,
as we can see in Eqs. (\ref{deltam2}) and (\ref{zeta2}) that $z_2/\Sigma(0)\sim 1/(e^2 \hat{\phi}^2)$.
We would also expect $z_4/z_2\sim 1/(e^2 \hat{\phi}^2)$.
So we would find that $|z_4 ~p^4| \sim|z_2 ~p^2|$ for $p^2\approx e^2 \hat{\phi}^2$ and the expansion in Eq. (\ref{expansion2}) may not give
the correct result.

A better way to evaluate the self-energy is to do expansion around $p^2=e^2 \hat{\phi}^2$.
Introducing
\begin{equation}
\delta {\tilde m}^{2}=- \Sigma(p^2=e^2\hat{\phi}^2),~~
{\tilde z}_{2}=-\left.\frac{\partial(\Sigma(p^2))}{\partial p^{2}}\right|_{p^2=e^2\hat{\phi}^2}, \label{expansion3}
\end{equation}
 $\Sigma(p^2)$ can be written as
\begin{equation}
	\begin{aligned}
		\Sigma \left(p^{2}\right)&=
		-\delta\tilde{m}^{2}-\tilde{z}_{2}(p^2-e^2\hat{ \phi}^{2})
		+\tilde{z}_{4}(p^2-e^2\hat{ \phi}^{2})^2+\cdots 
	\end{aligned}\label{expansion4}
\end{equation}
Again we would expect $\tilde{z}_{4}/\tilde{z}_{2}\sim 1/(e^2 \hat{\phi}^2)$, 
but the pole mass would deviate from the tree-level mass by an order $e^4$ correction and
we would expect $p^2-e^2\hat{ \phi}^{2} \sim e^4  \hat{\phi}^2$ in this case.
We would find  that the second term in Eq. (\ref{expansion4})  is expected to be order $e^6$ and
the third term in Eq. (\ref{expansion4}) is expected to be order $e^8$.  Expansion to higher order of $(p^2-e^2\hat{ \phi}^{2})^n$
would give corrections of order $e^{4+2n}$. This should be a valid expansion.
 
 Taking the first and the second terms in Eq. (\ref{expansion4}), we can rewrite the renormalized inverse propagator as
 \begin{equation}
	\begin{aligned}
		G_{\mu \nu}\left(p^{2}\right)^{-1}	&
		=ig_{\mu \nu}\left[p^2-e^{2} \hat{\phi}^{2}-\delta\tilde{m}^{2}-\tilde{z}_{2}(p^2-e^2\hat{ \phi}^{2})\right]+p_{\mu}p_{\nu}\,\mathrm{term}
	\end{aligned}.\label{green_function}
\end{equation}
The pole mass is found with the condition that the first term in  Eq. (\ref{green_function}) vanishes at $p^2=m_{v}^{2}$ .
This condition gives 
\begin{eqnarray}
m_{v}^{2}=e^2 \hat{\phi}^{2}+\delta\tilde{m}^{2}/(1-\tilde{z}_{2})
 \approx e^2 \hat{\phi}^{2}+\delta\tilde{m}^{2} \label{polemass2}
\end{eqnarray}
where we have neglected correction $ \tilde{z}_{2} \delta\tilde{m}^{2}$ which is of order $e^6$.
Some details of the calculation of the self-energy diagrams in this case are given in Appendix \ref{AppendixB}. 
We find 
\begin{equation}
	\begin{aligned}
		\delta\tilde{m}^{2 }=\frac{\hbar }{16 \pi^{2}} e^{4}\phi^{2} \left[ \xi-\frac{\xi}{2} \ln \left(\frac{e^2\hat{\phi}^2\xi m^2_2}{\mu^{4}}\right)  -\frac{62}{9}+\frac{10}{3}\ln \frac{e^{2} \hat{\phi}^{2}}{\mu^{2}}\right]  
	\end{aligned}.  \label{deltam2-a}
\end{equation}
The cancellation of the gauge dependence is similar to the case in Eq.(\ref{polemass1-a}). 
Evaluating the mass at the vacuum and plugging Eqs.  (\ref{vacuum}),  (\ref{v1_vacuum_final}) and 
(\ref{deltam2-a}) into Eq. (\ref{polemass2})  we get the pole mass at order $e^4$ as
\begin{equation}
	\begin{aligned}
		m_{v}^{2}=e^2\mu_{X}^{2}+\frac{\hbar}{16\pi^2}e^4\mu_{X}^{2}\left( -\frac{142}{9}+\frac{124}{9}\ln e^2 -\frac{40}{3} \ln^{2} e\right).
	\end{aligned} \label{polemass3}
\end{equation}
Eq. (\ref{polemass3}) is also a manifestly gauge invariant result, but it slightly differs from Eq. (\ref{polemass1-a}).

\section{Conclusion \label{sec:conc}}
\qquad
In summary, we have carefully calculated the one-loop self-energy contributions to the vector meson in CW model.
In our calculation, we use the dressed propagator which includes a resummation of the one-loop contribution to the
scalar propagators. These contributions are found to be of the same order of the tree-level scalar propagators
in the CW model and should be taken into account in a careful analysis. 
We evaluate the one-loop self-energy contribution as an expansion of momentum square both at around $p^2=0$ and 
at around $p^2=e^2 \hat{\phi}^2$.  The two results are slightly different, and we suggest that the latter one is the correct result
because it is obtained from a more reliable expansion.

We find that our results agree with the results obtained in  \cite{Kang:1974yj} if restoring to the tree-level scalar propagator.
However,  if taking the dressed propagator in calculation, the conclusion is very different from the conclusion obtained in \cite{Kang:1974yj}.
We find that the pole mass of the vector meson  calculated  using the dressed propagator is gauge invariant 
at the vacuum determined using the effective potential calculated with 
a similar resummation, i.e. a resummation of daisy diagrams.  

A related problem is the gauge invariance of  the scalar meson mass in the CW model.  
This problem will be discussed in another publication.

\section*{Acknowledgements}
\label{sec:acknowledgements}
LW would like to thank Han-qing Zheng for helpful discussion.
This work is supported by National Natural Science Foundation of China(NSFC), grant No. 11875130.

\appendix
\section{Dressed propagators and vertices} \label{AppendixA}
\qquad 
The dressed propagators we use are:
\begin{align}
	1~\fd{2cm}{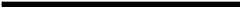}~1 ~~=~~ D_{11}(k) &= \frac{i}{k^2 - m_{1}^{2} } \label{D11def}\\
	2~\fd{2cm}{scalarline1}~2 ~~=~~ D_{22}(k) &=  \frac{i(k^2 - e^2 \xi \hat{\phi}^2)}{D(k)}\\
	\mu~\fd{2cm}{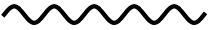}~\nu ~~=~~ \Delta_{\mu\nu}(k) &= -i \left[ \frac{1}{k^2 -e^2 \hat{\phi}^2}\left(g_{\mu\nu} - \frac{k_\mu k_\nu}{k^2} \right) + \frac{\xi (k^2 - m_{2}^{2})}{D(k)} \frac{k_\mu k_\nu}{k^2}\right]\\
	\mu~\fd{2cm}{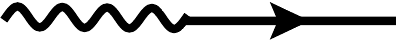}~2 ~~=~~ T_{\mu2} (k) &= \frac{\xi e \hat{\phi}}{D(k)} k_\mu \label{Tmu2def} 
\end{align}
where
\begin{equation}
	D(k)=(k^2-\beta_{1}^{2})(k^2-\beta_{2}^{2}), \qquad \beta_{1}^{2}\beta_{2}^2=e^2\hat{\phi}^2\xi m_{2}^{2},
	\qquad\beta_{1}^2+\beta_{2}^2=m^{2}_{2},
\end{equation}
and
\begin{equation}
	m^{2}_{1}(\hat{\phi})=\frac{\partial^2V_{e^{4}}}{\partial\hat{\phi}^2}=\frac{\lambda}{2} \hat{\phi}^{2}+\frac{\hbar e^{4}}{16 \pi^{2}} \hat{\phi}^{2}\left(3+18 \ln \frac{e \hat{\phi}}{\mu}\right),
\end{equation}

\begin{equation}
	m^{2}_{2}(\hat{\phi})=\frac{1}{\hat{\phi}}\frac{\partial V_{e^{4}}}{\partial \hat{\phi}}=\frac{\lambda}{6}\hat{\phi}^{2}+\frac{\hbar e^4}{16 \pi^2}\hat{\phi}^{2}(-1+6 \ln\frac{e\hat{\phi}}{\mu}).
\end{equation}

The effective vertices under a background field $\phi_1=\hat{\phi}$ are as follows.
\be
\hspace{-2mm}
\resizebox{10mm}{!}{
	\parbox{10mm}{
		\begin{tikzpicture}[]
			\node (label) at (0,0){ \fd{1.5cm}{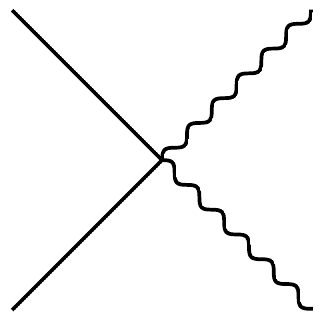} };
			\node[rotate=0,anchor=east] at (-0.65,0.7) {$i$};
			\node[rotate=0,anchor=east] at (-0.65,-0.5) {$j$};
			\node[rotate=0,anchor=east] at (1.15,-0.5) {$\nu$};
			\node[rotate=0,anchor=east] at (1.15, 0.7) {$\mu$};
		\end{tikzpicture}
}}
\hspace{8mm}
= 2i e^2 \delta_{ij}g^{\mu\nu}
\hspace{8mm}
\resizebox{10mm}{!}{
	\parbox{10mm}{
		\begin{tikzpicture}[]
			\node (label) at (0,0){ \fd{1.5cm}{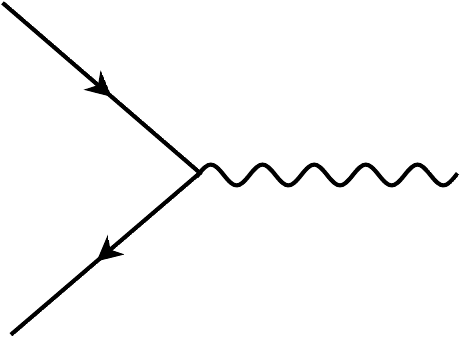} };
			\node[rotate=0,anchor=east] at (-0.65,0.7) {$i$};
			\node[rotate=0,anchor=east] at (-0.65,-0.5) {$j$};
			\node[rotate=0,anchor=east] at (0.9,0.3) {$\mu$};
		\end{tikzpicture}
}}
\hspace{8mm}
= e\epsilon_{ij}(k_i+k_j)^{\mu}
\hspace{8mm}
\resizebox{10mm}{!}{
	\parbox{10mm}{
		\begin{tikzpicture}[]
			\node (label) at (0,0){ \fd{1.5cm}{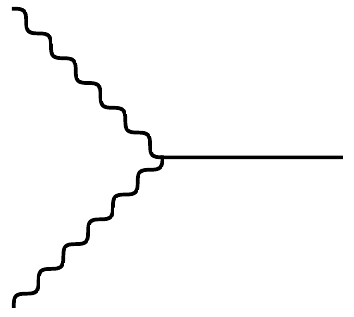} };
			\node[rotate=0,anchor=east] at (-0.65,0.7) {$\mu$};
			\node[rotate=0,anchor=east] at (-0.65,-0.5) {$\nu$};
			\node[rotate=0,anchor=east] at (0.9,0.3) {$j$};
		\end{tikzpicture}
}}
\hspace{10mm}
= 2ie^2\hat{\phi} \delta_{j1}g^{\mu\nu}
\ee
\be
\hspace{-2mm}
\resizebox{10mm}{!}{
	\parbox{10mm}{
		\begin{tikzpicture}[]
			\node (label) at (0,0){ \fd{1.5cm}{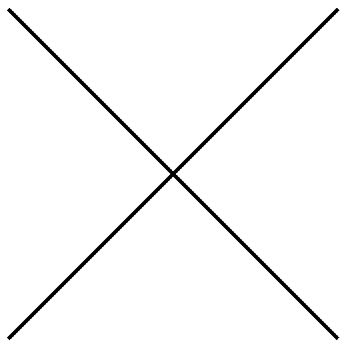} };
			\node[rotate=0,anchor=east] at (-0.65,0.7) {$i$};
			\node[rotate=0,anchor=east] at (-0.65,-0.5) {$j$};
			\node[rotate=0,anchor=east] at (1.15,-0.5) {$k$};
			\node[rotate=0,anchor=east] at (1.15, 0.7) {$l$};
		\end{tikzpicture}
}}
\hspace{9mm}
= -i\frac{\lambda}{3}(\delta_{ij}\delta_{kl}+\delta_{ik}\delta_{jl}+\delta_{il}\delta_{jk})\
\hspace{8mm}
\resizebox{10mm}{!}{
	\parbox{10mm}{
		\begin{tikzpicture}[]
			\node (label) at (0,0){ \fd{1.5cm}{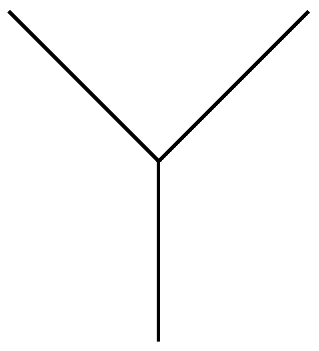} };
			\node[rotate=0,anchor=east] at (-0.65,0.7) {$i$};
			\node[rotate=0,anchor=east] at (1.15, 0.7) {$j$};
			\node[rotate=0,anchor=east] at (0,-0.5) {$k$};
		\end{tikzpicture}
}}
\hspace{9mm}
= - i \frac{\lambda}{3}\hat{\phi}(\delta_{ij}\delta_{k1}+\delta_{jk}\delta_{i1}+\delta_{ki}\delta_{j1})
\ee

\section{Evaluation of one-loop self-energy diagrams } \label{AppendixB}
\begin{figure}[H]
	\centering
	\includegraphics[width=0.9\linewidth]{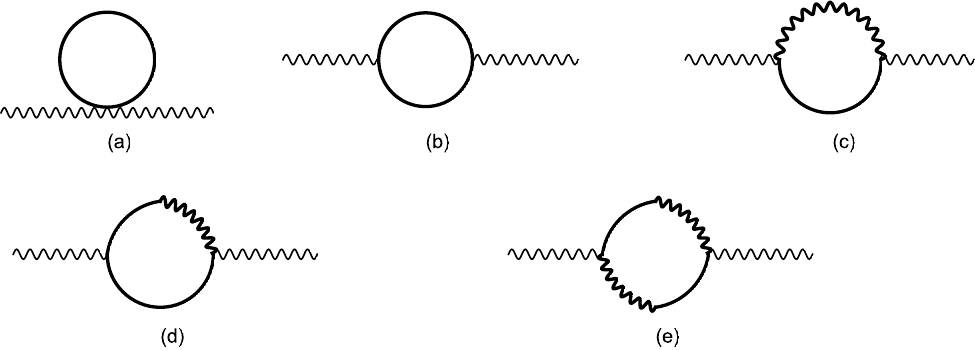}
	\caption{One-loop self-energy diagrams for vector meson}
	\label{fig:vector}
\end{figure}

The corresponding Feynman integrals are
\begin{eqnarray}
-i\Sigma_{\mu\nu}^{(a)}(p^2)&&=\frac{1}{2}\int \frac{d^Dk}{(2 \pi)^{D}} 2 i e^{2}  g_{\mu \nu}\left[D_{11}(k)+D_{22}(k)\right] \nonumber\\
	&&=\frac{1}{2}\int \frac{d^Dk}{(2 \pi)^{D}} 2 i e^{2}  g_{\mu\nu}\left[\frac{i}{k^{2}-m_{1}^{2}}+\frac{i\left(k^{2}-e^{2} \xi\hat{ \phi}^{2}\right)}{D(k)}\right], \label{B1-a} \\
      -i\Sigma_{\mu\nu}^{(b)}(p^2)&&=\int \frac{d^{D} k}{(2 \pi)^{D}} e \varepsilon_{12}(2 k+p)_{\mu} D_{11}(p+k)e \varepsilon_{2 1}(2 k+p)_{\nu} D_{22}(k)  \nonumber\\
 &&=4e^2\int \frac{d^{D} k}{(2 \pi)^{D}} \frac{k_{u} k_{\nu}\left(k^{2}-e^{2} \xi \hat{\phi}^{2}\right)}{\left((p+k)^{2}-m_{1}^{2}\right) D(k)}+p_{\mu}p_{\nu}\,\mathrm{term}, \label{B1-b}\\		
	-i\Sigma_{\mu\nu}^{(c)}(p^2)&&=\int \frac{d^{D} k}{(2 \pi)^{D}}2 i e^{2} \hat{\phi} g_{\mu \alpha} D_{11}(p+k)2 i e^{2} \hat{\phi} g_{\beta \nu} \Delta^{\alpha\beta}(k)   \nonumber\\
	&&=-4e^4\hat{ \phi}^2\int \frac{d^{D} k}{(2 \pi)^{D}}\left[ \frac{g_ {\mu \nu}}{\left((p+k)^{2}-m_{1}^{2}\right)\left(k^{2}-e^{2} \hat{\phi}^{2}\right)}+\frac{\xi k_{u} k_{\nu}}{\left((p+k)^{2}-m_{1}^{2}\right) D(k)}\right.  \nonumber\\
		&&\left. -\frac{k_{u} k_{\nu}}{\left((p+k)^2-m_{1}^{2}\right)\left(k^{2}-e^{2} \hat{\phi}^{2}\right) k^{2}}-\frac{\xi m_{2}^{2} k_{\mu}k_{\nu}}{\left((p+k)^{2}-m_{1}^{2}\right) D(k)k^2}\right], \label{B1-c}\\
	-i\Sigma_{\mu\nu}^{(d)}(p^2)&&=4\int \frac{d^{D} k}{(2 \pi)^{D}}e \varepsilon_{21}(2 p+k)_{\mu}  T_{\rho 2} (k) ~2 i e^{2} \hat{\phi} g^\rho_{~\nu} D_{11}(p+k) \nonumber\\
	&&=4\int \frac{d^{D} k}{(2 \pi)^{D}}  \frac{2 e^{4} \hat{\phi}^{2} \xi k_{\mu} k_{\nu}}{((p+k)^{2}-m_{1}^{2})D(k)}+p_{\mu}p_{\nu}\,\mathrm{term}, \label{B1-d}\\
		-i\Sigma_{\mu\nu}^{(e)}(p^2)&&=0. \label{B1-e}
\end{eqnarray}
In these equations, the space-time dimension $D$ is $D=4-2\varepsilon$. 
Notice that contributions from the second term in the parenthesis in Eq. (\ref{B1-b}), the second term in  Eq. (\ref{B1-c})
and the first term in Eq.  (\ref{B1-d}) add up to zero.

In order to simplify the calculation, we add all the above terms and get
\begin{equation}
	\begin{aligned}
		-i\Sigma_{\mu\nu}&(p^2)=-i\Sigma_{\mu\nu}^{(a)}(p^2)+4e^2\int \frac{d^{D} k}{(2 \pi)^{D}} \frac{k_{u} k_{\nu}k^{2}}{\left((p+k)^{2}-m_{1}^{2}\right) D(k)}\\	&-4e^4\hat{ \phi}^2\int \frac{d^{D} k}{(2 \pi)^{D}}\left[ \frac{g_ {\mu \nu}}{\left((p+k)^{2}-m_{1}^{2}\right)\left(k^{2}-e^{2} \hat{\phi}^{2}\right)}-\frac{k_{u} k_{\nu}}{\left((p+k)^2-m_{1}^{2}\right)\left(k^{2}-e^{2} \hat{\phi}^{2}\right) k^{2}}\right. \\
		&\left. -\frac{\xi m_{2}^{2} k_{\mu}k_{\nu}}{\left((p+k)^{2}-m_{1}^{2}\right) D(k) k^{2}}\right] +p_{\mu}p_{\nu}\,\mathrm{term}
	\end{aligned}\label{general_fey}
\end{equation}

After Feynman parameterization, we separate the remaining expression in Eq. (\ref{general_fey}) into five parts and obtain
\begin{equation}
	-i\Sigma_{\mu\nu}(p^2)=-i\Sigma^{(a)}_{\mu\nu}(p^2)+\left( \mathrm{\RNum{1}}\right) +\left( \mathrm{\RNum{2}}\right) +\left( \mathrm{\RNum{3}}\right)+\left( \mathrm{\RNum{4}}\right) \label{integral}
\end{equation}
where $\Sigma^{(a)}_{\mu \nu}$ is independent of $p^2$ and is given in Eq. (\ref{C1-a}) in Appendix \ref{AppendixC}. 
Other terms are 
\begin{eqnarray}
		\left( \mathrm{\RNum{1}}\right) &&=\frac{i g_{\mu\nu}}{16 \pi^{2}}\left(2 e^{4}\hat{\phi}^{2}\right) \int_{0}^{1} d x 
		\ln \left(\frac{M^{2}(a_{1},b_{1},c_{1})}{\mu^{2}}\right)M^{2}\left( -\frac{a_{1}}{c_{1}},-\frac{b_{1}}{c_{1}},1\right), \label{B2-1}\\
		\left( \mathrm{\RNum{2}}\right)&&=
		\frac{i g_{\mu \nu}}{16 \pi^{2}} e^{2}   \int_{0}^{1}  d x ~\frac{1}{d} \left.\Big[ \ln \left( M^{2}(a,b,c)\right) M^{2}\left(a,b,c\right) M^{2}\left( 5a,3b,5c+2d\right) \right.  \nonumber\\
		&&\left.- \ln \left( M^{2}(a,b-d,c+d)\right)  M^{2}\left(a,b-d,c+d\right)  M^{2}\left( 5a,3b-3d,5c+3d\right) \Big] \right., \label{B2-2}\\
		\left( \mathrm{\RNum{3}}\right)&&=\frac{i g_{\mu \nu}}{16 \pi^{2}} e^{2} \left[ \int_{0}^{1} d x_{1} \int_{0}^{1-x_{1}} d x_{2} ~
		4  \widehat{M}^{2}+\int_{0}^{1} d x(2 e^{2} \hat{\phi}^{2})(1-x)\right.  \nonumber\\
		&&\left.-\int_{0}^{1} dx \left(\frac{3}{2 d}\left( \left(M^{2}(a,b,c)\right)^2-\left(M^{2}(a,b-d,c+d)\right)^2 \right)-2 ax^{2}(1-x) \right)
		\right], \label{B2-3} ~~\\
		\left( \mathrm{\RNum{4}}\right)&&=\frac{i g_{\mu \nu}}{16 \pi^{2}} e^{2} \Delta_{\varepsilon} 
		\left[ -4 e^{2} \hat{\phi}^{2}+\int_{0}^{1} d x_{1}  \int_{0}^{1-x_{1}} d x_{2}
		\left(6 \widehat{M}^{2}+2 l^{2}+2e^2\hat{\phi}^2 \right)
		 \right]  ,\label{B2-4}
\end{eqnarray}
where 
\begin{equation}
	\begin{aligned}
		&M^{2}(a,b,c)=ax^2+bx+c,\qquad \widehat{M}^{2} = ax_{1}^{2}+bx_{1}+c+dx_{2},\qquad l^2=ax_1^2,\\
		&a=p^2,\qquad b=m_{1}^{2}-\beta_{2}^{2}-p^{2},\qquad c=\beta_{2}^{2},\qquad d=\beta_{1}^{2}-\beta_{2}^{2},\\
		&a_{1}=p^2, \qquad b_{1}=m_{1}^{2}-e^2\hat{ \phi}^{2}-p^{2},\qquad c_{1}=e^2\hat{\phi}^{2},
	\end{aligned}
\end{equation}
and
\begin{equation}
	\Delta_{\varepsilon}=\frac{1}{\varepsilon}-\gamma_{E}+\ln 4\pi.
\end{equation}
In Eqs. (\ref{integral}), (\ref{B2-1}),  (\ref{B2-2}), (\ref{B2-3}) and  (\ref{B2-4}),  we have neglected all terms proportional to $p_\mu p_\nu$.
So far we have not made approximation in the evaluation of self-energy diagrams except neglecting terms proportional to $p_\mu p_\nu$.
A further examination shows that the last term in Eq. (\ref{B1-c}) gives higher order contribution which can be neglected in our analysis. 
If neglecting this contribution, Eqs. (\ref{B2-1}) and (\ref{B2-2})  would be re-written as other forms.

Notice that two terms in the bracket in Eq. (\ref{B2-2}) cancel with each other when $d=0$. When $d$ is small compared to other parameters, 
these two terms would give a result proportional to a factor $d$  which cancels the $d$ in the denominator outside the bracket.
This is the case for $p^2\approx e^2 \hat{\phi}^2$. A careful calculation should be done to extract result in this case.
After a careful calculation,
an expression for the one-loop contribution to the self-energy of vector meson is found at $p^2\approx e^2 \hat{\phi}^2$ as
\begin{equation}
\begin{aligned}
			-i \Sigma_{\mu \nu}\left(p^{2}\right)=&\frac{i\hbar g_{\mu \nu}}{16 \pi^{2}}\left[e^{4} \phi^{2}\left( \xi-\frac{\xi}{2} \ln \left(\frac{e^{2} \hat{\phi}^{2} \xi m_{2}^{2}}{\mu^{4}}\right) -\frac{62}{9}+\frac{10}{3} \ln \frac{e^{2} \hat{\phi}^{2}}{\mu^{2}}\right ) \right. \\
			 &\left.+\left(\frac{31}{9}-\frac{5}{3} \ln \frac{e \hat{\phi}^{2}}{\mu^{2}}+2 \ln \frac{m_{1}^{2}}{\mu^{2}}\right)e^2(p^2-e^2\hat{ \phi}^2 )\right] ,
	\end{aligned}  \label{self-energy1}
\end{equation}
where we have eliminated terms proportional to $\Delta_{\varepsilon}$ by subtraction and neglected  terms proportional to $p_\mu p_\nu$.
We have also restored a factor $\hbar$ in Eq. (\ref{self-energy1}) for one-loop correction.
Using Eq. (\ref{self-energy1}), we can find
\begin{equation}
\delta \tilde{m}^{2}=\frac{\hbar}{16 \pi^{2}} e^{4} \phi^{2}\left[\xi-\frac{\xi}{2} \ln \left(\frac{e^{2} \hat{\phi}^{2} \xi m_{2}^{2}}{\mu^{4}}\right)-\frac{62}{9}+\frac{10}{3} \ln \frac{e^{2} \hat{\phi}^{2}}{\mu^{2}}\right],
\end{equation}
and
 \begin{equation}
\tilde{z}_{2}=\frac{\hbar e^{2}}{16 \pi^{2}}\left(\frac{31}{9}-\frac{5}{3}\ln \frac{e\hat{ \phi}^2}{\mu^2}+2\ln \frac{m_{1}^{2}}{\mu^2}\right).
\end{equation}

We note that when evaluating the self-energy diagrams in  Fig. \ref{fig:vector} at $p^2=e^2 \hat{\phi}^2$,
we would find imaginary parts in some of the diagrams.
There is an imaginary part in Fig. \ref{fig:vector}(b) which does not depend on the gauge-fixing parameter $\xi$.
At the order we are considering, this imaginary part is cancelled by a contribution from Fig. \ref{fig:vector}(c) .
Other contributions to the imaginary parts are proportional to the gauge-fixing parameter $\xi$.
These terms add up to zero and the imaginary parts cancel completely.
So we do not have imaginary part in the final result of the self-energy of vector meson.

\section{Summary of one-loop self-energy contributions} \label{AppendixC}
The self-energy contributions given in Eqs. (\ref{B1-a}),  (\ref{B1-b}),  (\ref{B1-c}),  (\ref{B1-d}) and  (\ref{B1-e}) can be evaluated as
an expansion of $p^2$ around $p^2=0$.  The results to order $p^2$  are
\begin{eqnarray}
		-i\Sigma_{\mu\nu}^{(a)}(p^2)&&=\frac{i g_{\mu\nu}}{16 \pi^{2}}\left[e^{4} \xi \phi^{2}+e^{4}\xi \phi^{2}\left(\Delta_{\varepsilon}-\frac{1}{2} \ln \left(\frac{e^{2} \hat{\phi}^{2} \xi m_{2}^{2}}{\mu^{4}}\right)\right) \right], \label{C1-a}\\
		-i\Sigma_{\mu\nu}^{(b)}(p^2)&&=\frac{i g_{\mu\nu}}{16\pi^{2}}\left[-\frac{3}{2}e^4\hat{\phi}^{2}\xi+\left( -\frac{5}{18}+\frac{2m_{1}^{2}}{3m_{2}^{2}}\right)e^2p^2 \right. \nonumber \\
		&&\left. -( e^4\hat{ \phi}^{2}\xi+\frac{1}{3}e^2p^2)\left(\Delta_{\varepsilon}-\frac{1}{2}\ln \left(\frac{e^{2} \hat{\phi}^{2} \xi m_{2}^{2}}{\mu^{4}}\right) \right) \right], \label{C1-b}\\
		-i\Sigma_{\mu\nu}^{(c)}(p^2)&&=\frac{i g_{\mu\nu}}{16\pi^{2}}\left[-(\frac{5}{2}+\frac{3}{2}\xi)e^4\hat{ \phi}^2-2\left( 1-\frac{m_{1}^{2}}{3m_{2}^{2}}-\frac{1}{6}\ln\frac{e^2\hat{ \phi}^2}{\mu^2}
		+\frac{1}{12}\ln \left(\frac{e^{2} \hat{\phi}^{2} \xi m_{2}^{2}}{\mu^4}\right)\right)e^2 p^2\right. \nonumber\\ 
		&& \left.-(3+\xi)e^4\hat{ \phi}^2\Delta_{\varepsilon} +e^4\hat{ \phi}^2\left( 3\ln\frac{e^2\hat{ \phi}^2}{\mu^{2}}+\frac{1}{2}\xi\ln \left(\frac{e^{2} \hat{\phi}^{2} \xi m_{2}^{2}}{\mu^{4}}\right)\right) \right], \label{C1-c}\\
		-i\Sigma_{\mu\nu}^{(d)}(p^2)&&= \frac{i g_{\mu\nu}}{16\pi^{2}}\left[ 3e^4\hat{ \phi}^2\xi-\frac{4m_{1}^{2}}{3m_{2}^{2}}e^2p^2
		+2e^4\hat{ \phi}^{2}\xi\left(\Delta_{\varepsilon}-\frac{1}{2}\ln \left(\frac{e^{2} \hat{\phi}^{2} \xi m_{2}^{2}}{\mu^{4}}\right) \right) \right], \label{C1-d}\\
		-i\Sigma_{\mu\nu}^{(e)}(p^2)&&=0. \label{C1-e}
\end{eqnarray}
In these results,  we have neglected all terms proportional to $p_\mu p_\nu$.
We note that  $-i\Sigma_{\mu\nu}^{(a)}$ is independent of $p^2$, 
so we  have $-i\Sigma_{\mu\nu}^{(a)}(0)=-i\Sigma_{\mu\nu}^{(a)}(e^2\hat{\phi}^{2})$.

Summarizing results in Eqs. (\ref{C1-a}),  (\ref{C1-b}),  (\ref{C1-c}),  (\ref{C1-d}) and  (\ref{C1-e}),
we find the one-loop contribution to the self-energy as an expansion at $p^2=0$ as
\begin{equation}
\begin{aligned}
-i \Sigma_{\mu \nu}\left(p^{2}\right) &= \frac{i\hbar g_{\mu \nu}}{16 \pi^{2}}\left[ e^{4} \phi^{2}\left( \xi-\frac{5}{2}-\frac{\xi}{2} \ln \left(\frac{e^{2} \hat{\phi}^{2} \xi m_{2}^{2}}{\mu^{4}}\right)+3 \ln \frac{e^{2} \hat{\phi}^{2}}{\mu^{2}}\right) \right. \\
 &\left. +\left(-\frac{41}{18}+\frac{1}{3} \ln \frac{e^{2} \hat{\phi}^{2}}{\mu^{2}}\right)e^2p^2 \right] ,\label{self-energy4}
 \end{aligned}
\end{equation}
where we have eliminated terms proportional to $\Delta_{\varepsilon}$ by subtraction and neglected  terms proportional to $p_\mu p_\nu$.
We have also restored a factor $\hbar$ in Eq. (\ref{self-energy4}) for one-loop correction.

If we do not use the dressed propagator of scalars and restore to the tree-level propagator, 
we just need to take $m_{1}^{2}=\frac{\lambda}{2}\hat{\phi}^2$ and $m_{2}^{2}=\frac{\lambda}{6}\hat{\phi}^2$.
If taking the tree-level masses of the scalar mesons and neglecting the coupling constants and the gauge-fixing parameter $\xi$ in logarithms, 
 our results agree with the results in ~\cite{Kang:1974yj}.

For $p^2\approx e^2 \phi^2$, we can find the self-energy contributions as follows.
\begin{eqnarray}
-i\Sigma_{\mu\nu}^{(a)}(p^2)&&=\frac{i g_{\mu\nu}}{16 \pi^{2}}\left[e^{4} \xi \phi^{2}+e^{4}\xi \phi^{2}\left(\Delta_{\varepsilon}-\frac{1}{2} \ln \left(\frac{e^{2} \hat{\phi}^{2} \xi m_{2}^{2}}{\mu^{4}}\right)\right) \right], \label{C2-1}\\
		-i\Sigma_{\mu\nu}^{(b)}(p^2)&&=\frac{ig_{\mu\nu}}{16\pi^{2}}\left[-\frac{e^2p^2}{3}\Delta_{\varepsilon}-\frac{8}{9}e^2p^2+\frac{1}{3}e^2p^2\ln\frac{p^2}{\mu^2}+\frac{1}{3}i\pi e^2p^2 \right. \nonumber \\
		&& \left. -e^4\xi\hat{\phi}^2\left( \Delta_{\varepsilon}-2-\ln\frac{p^2}{\mu^2}-i\pi\right) \right],
		\label{C2-2}\\
		-i\Sigma_{\mu\nu}^{(c)}(p^2)&&=\left( \mathrm{\RNum{1}}\right)+\frac{ig_{\mu\nu}}{16\pi^{2}}\left[e^4\hat{ \phi}^2(1-3\Delta_{\varepsilon})+\frac{5}{9}e^2p^2-\frac{1}{3}e^2p^2\ln\frac{p^2}{\mu^2}-\frac{1}{3}i\pi e^2p^2 \right. \nonumber \\
		&&\left. -e^4\xi\hat{\phi}^2\left( \Delta_{\varepsilon}-2-\ln\frac{p^2}{\mu^2}-i\pi\right)\right], \label{C2-3}\\
		-i\Sigma_{\mu\nu}^{(d)}(p^2)&&=\frac{ig_{\mu\nu}}{16\pi^{2}}\left[e^4\xi\hat{\phi}^2\left( 2\Delta_{\varepsilon}-4-2\ln\frac{p^2}{\mu^2}-2i\pi\right)  \right],\label{C2-4} \\
		-i\Sigma_{\mu\nu}^{(e)}(p^2)&&=0. \label{C2-5}
\end{eqnarray}
where 
\begin{eqnarray}
		\left( \mathrm{\RNum{1}}\right)&&=
		\frac{i g_{\mu \nu}}{16 \pi^{2}}\left( e^{4} \hat{\phi}^{2}\right)\left[\left( -\frac{20}{3}-\frac{5 p^{2}}{9 e^{2} \hat{\phi}^{2}}-\frac{e^{2} \hat{\phi}^{2}}{3 p^{2}}\right)+\left(\frac{3}{2}+\frac{p^{2}}{6e^{2} \hat{\phi}^{2}}+\frac{3 e^{2} \phi^{2}}{2 p^{2}}+\frac{e^{4} \hat{\phi}^{4}}{6 p^{4}}\right) \ln \left(\frac{e^{2} \hat{\phi}^{2}}{\mu^{2}}\right)\right. \nonumber \\ 
		&&\left.+\left(\frac{3}{2}+\frac{p^{2}}{6 e^{2} \hat{\phi}^{2}}-\frac{3 e^{2} \phi^{2}}{2 p^{2}}-\frac{e^{4}\hat{\phi}^{4}}{6 p^{4}}\right) \ln \left(\frac{m_{1}^{2}}{\mu^{2}}\right) \right.  \nonumber \\
		&& \left.+\left(-\frac{5}{3}-\frac{p^{2}}{6 e^{2} \hat{\phi}^{2}}-\frac{e^{2} \hat{\phi}^{2}}{6 p^{2}}\right)\left(x_{+}-x_{-}\right) \ln \left(\frac{x_{+}+x_{+} x_{-}}{x_{-}+x_{+} x_{-}}\right)\right]  
\end{eqnarray}
with 
\begin{equation}
	x_{+}+x_{-}=\frac{m_{1}^{2}-e^2\hat{ \phi}^{2}-p^{2}}{p^2},\qquad x_{-} x_{+}=\frac{e^2\phi^2}{p^2}.
\end{equation}
Again, we have neglected all terms proportional to $p_\mu p_\nu$ in Eqs. (\ref{C2-2}),  (\ref{C2-3}) and  (\ref{C2-4}).
Using Eqs.  (\ref{C2-1}),  (\ref{C2-2}),  (\ref{C2-3}) and  (\ref{C2-4}), we can obtain $\Sigma_{\mu \nu}$ as an expansion around $p^2=e^2\phi^2$
and get Eq.  (\ref{self-energy1}).

\providecommand{\href}[2]{#2}\begingroup\raggedright
\endgroup

\bibliography{paper}

\bibliographystyle{utphys}

\end{document}